\documentclass[12pt]{article}

\title{\bf A topological signature in cosmic topology}
\author{G.I. Gomero\thanks{E-mail: german@cbpf.br} , \  
M.J. Rebou\c{c}as\thanks{E-mail: reboucas@cbpf.br} , \ 
A.F.F. Teixeira\thanks{E-mail: teixeira@cbpf.br} , \ 
\\ 
Centro Brasileiro de Pesquisas F\'\i sicas, \\
Rua Dr. Xavier Sigaud 150, \\
22290-180 Rio de Janeiro -- RJ, Brazil
}

\begin{document}

\date{}
\maketitle


\vspace{5mm}
\begin{abstract} 
\begin{sloppypar}
Two procedures for obtaining (extracting and constructing) the topological 
signature of any multiply connected Robertson-Walker (RW) universe 
are presented. It is shown through computer-aided simulations that 
both approaches give rise to the same topological signature for a 
multiply connected f\/lat RW universe. The strength of these approaches 
is  illustrated by  extracting the topological signatures of a f\/lat 
($k=0$), an elliptic ($k=1$), and a hyperbolic ($k=-1$) multiply 
connected RW universes. We also show how separated contributions 
of the covering isometries add up to form the topological signature 
of a RW f\/lat universe. There emerges from our theoretical 
results and simulations that the topological signature arises 
(in the mean) even when there are just a few images for each object.
It is also shown that the mean pair separation histogram 
technique works, and that it is a suitable approach for 
studying the topological signatures of RW universes as well as 
the role of non-translational isometries.
\end{sloppypar}
\end{abstract}     

\newpage
\section{Introduction}
\label{intro}
\setcounter{equation}{0}

The astrophysical observations indicate that to a high degree of
accuracy our universe is locally homogeneous and isotropic. Thus
in the framework of the general relativity theory it can be 
described through a Robertson-Walker (RW) metric
\begin{equation}
\label{RWmetric}
ds^2 = dt^2 - R^2(t)\, d \sigma^2 \; ,
\end{equation}
where $t$ is a cosmic time, and 
$d\sigma^2 = d\chi^2 + f^2(\chi)\,[\,d\theta^2+\sin^2\theta\,d\phi^2\,]$ 
with $f(\chi)= \chi\,,\;$ $\sin\chi\,,\;$ $\sinh\chi\,,\;$ depending on 
the sign of the constant spatial curvature ($k = 0, \pm 1$). Clearly  
$R(t)$ is the scale factor that carries the unit of length. 

However, whether we live in a simply or multiply connected space and 
what is the size and the shape of the universe are open problems, 
which have received increasing attention over the past few years
(see, for example, \cite{LaLu}~--~\cite{Weeks01} and references therein).
The most immediate consequence of multiply-connectedness of the 
universe is the possibility of observing multiple images of 
cosmic objects, whose existence can be perceived by the
simple reasoning presented below. 

It is often assumed that the $t=const$ spatial sections $M$ of 
a RW spacetime manifold are one of the following simply 
connected spaces:  Euclidean $E^3$ ($k=0$), elliptic $S^3$ ($k=1$), 
or the hyperbolic $H^3$ ($k=-1$), depending on the sign of the 
constant spatial curvature $k$.
However, one can lift the restriction of simply-connectedness 
for the spatial sections  by allowing $M$ to be any one of the
possible quotient manifolds $M = \widetilde{M}/\Gamma$, where 
$\widetilde{M}$ stands for $E^3$, $S^3$ or $H^3$ (depending on
the sign of $k$), and $\Gamma$ is a discrete subgroup of the 
full group of isometries of the covering manifold $\widetilde{M}$ 
acting freely on $\widetilde{M}$~\cite{Wolf}.
The action of $\Gamma$ tesselates $\widetilde{M}$ into identical 
cells or domains which are copies of what is known as fundamental 
polyhedron (FP). In forming the quotient manifolds $M$ the 
essential point is that they are obtained from $\widetilde{M}$ 
by identifying points which are equivalent under the action of 
the discrete group $\Gamma$. Hence, each point on the quotient 
manifold $M$ represents all the equivalent points on the 
covering manifold $\widetilde{M}$.
On the other hand, in the general relativity approach to cosmological 
modelling, the physicists  assume that our universe can be modelled 
by a manifold, and thus a given cosmic object is described  by a 
point $p \in M$, which represents, when $M$ is multiply connected,  
a set of equivalent points (images of $p$) on the covering manifold
$\widetilde{M}$. 
So, to f\/igure out that multiple images of an object can indeed be 
observed if the universe is multiply-connected, consider that the 
observed universe is a ball $\mathcal{B}_{R_H} \subset \widetilde{M}$ whose 
radius $R_H$ is the particle horizon, and denote by $L$ the largest lenght 
of the fundamental polyhedron $P$ of $M$ ($P \subset \widetilde{M}$). 
Thus, when $R_H > L/2$, for example, the set of (multiple) images of an 
given object that lie in $\mathcal{B}_{R_H}$ can in principle be observed.
Obviously the observable images of an object constitute a f\/inite 
subset of the set of all equivalent images of the object.%
\footnote{For the precise and general condition for the existence 
of multiple images see~\cite{GTRB98}. Also note that for
$R_H > 2\,r_{inj}$ (here $r_{inj}$ is the injectivity radius, see 
e.g.~\cite{LuRou} for notation) multiple images can still arise.}

General relativity (GR) relates the matter content of the
universe to its geometry, and reciprocally the geometry constrains 
the dynamics of the matter content. As GR is a purely metrical
(local) theory it cannot be used (without additional topological 
assumptions as, e.g., asymptotical f\/latness or asymptotical locally 
anti-de Sitter~\cite{Galloway99}) to settle the global structure 
(topology) of spacetime. 
One way to tackle the problems regarding the topology of the 
universe is through a suitable statistical analysis applied to 
catalogues of discrete cosmic sources to f\/ind out whether there 
are multiple images of cosmic objects, and eventually determine 
the shape and the size of the universe from the pattern of the 
repeated images.

Cosmic crystallography~\cite{LeLaLu} (CC) is one of such statistical 
methods, which looks for distance correlations between the images 
of cosmic objects using pair separation histograms (PSH), i.e. 
graphs of the number of pairs of sources versus the 
distance between them. These correlations arise from 
the discrete isometries of the covering group $\Gamma$, which 
give rise to the (observed) multiple images.
The initial expectation behind the crystallographic method 
was that these distance correlations would manifest as 
sharp peaks (called spikes) in PSH's, and that the
spike spectrum (their positions and relative amplitudes) 
would be a def\/inite signature of the topology.
The f\/irst simulations performed for some specif\/ic f\/lat manifolds
appeared to conf\/irm these expectations~\cite{LeLaLu} in 
that the PSH's corresponding to those particular manifolds 
presented distinct spike spectra.
Afterwards, however, histograms were generated for the specif\/ic 
cases of Weeks~\cite{LeLuUz} and one of the Best~\cite{FagGaus} 
manifolds and no spikes were found. Their graphs show that,
within the degree of accuracy of the corresponding plots, 
the PSH's of these hyperbolic manifolds exhibit no spikes.
However, the spikes in PSH's can either be of topological origin 
or arise from purely statistical f\/luctuations. 
So, it is important to consider that the statistical noise 
in PSH's can both give rise to sharp peaks of purely statistical
origin, and even hide the topological spikes, which are the
sharp peaks that really matter in cosmic crystallography.
Thus, the ultimate proof (or disproof) for the existence 
of the topological spikes in a class of manifolds cannot rely
only on graphs, since they do not constitute a proof but a
clue or simply an indication.%
\footnote{For preciseness, it should be added that in~\cite{LeLuUz} 
two causes are presented for the absence of spikes in the simulated 
PSH corresponding to the Weeks manifold. Note, however, that the 
explanation in~\cite{LeLuUz} for the lack of spikes in PSH's does 
not seem to match with those presented in 
ref.~\cite{GTRB98}.}
As a matter of fact, the proof for the existence of topological 
spikes ought to arise from a theoretical statistical analysis 
of the distance correlations to reveal the role in PSH's played 
by all types of isometries of the covering group $\Gamma$.

In considering discrete astrophysical sources in the context of 
multiply connected RW spacetimes, the \emph{observable universe} 
is the region or part of the universal covering 
manifold $\widetilde{M}\,$  causally connected to an image 
of a given observer since the moment of matter-radiation 
decoupling. Clearly in the observable universe one has the
set of observable images of the cosmic objects, denoted by
$\mathcal{O}\,$. A catalogue is a particular subset 
$\mathcal{C} \subset \mathcal{O}$, of \emph{observed} images, 
since by several observational limitations one can hardly 
record all the images present in the observable universe. 
Our observational limitations can be formulated through 
selection rules which dictate how the subset $\mathcal{C}$ 
arises from $\mathcal{O}$. 
Catalogues whose images obey the same well-behaved distribution
and that follow the same selection rules are said to be comparable 
catalogues (for more details about the formalization of these 
concepts see~\cite{GTRB98}).

In a recent article, in the context of Robertson-Walker spacetimes,
Gomero \emph{et al.\/}~\cite{GTRB98} have derived an expression for 
the expected pair separation histogram (EPSH) for an ensemble of 
comparable catalogues $\mathcal{C}$ with approximately the same 
number of sources and corresponding to a manifold $M$.
{}From the expression for the EPSH they have shown that the
spikes of topological nature in PSH's are due to Clif\/ford
translations%
\footnote{A Clif\/ford translation is an isometry $g$ of $\widetilde{M}$ 
such that for all $p \in \widetilde{M}$, the distance $d(p,gp)$ is 
constant.} 
whereas the other non-translational isometries
manifest as slight deformations of the EPSH corresponding to
the underlying simply connected manifold.
This general result holds regardless of the 3-geometry of the 
spacelike sections of the RW spacetimes, and its restriction 
to the specif\/ic case of Euclidean and hyperbolic 3-geometries
gives rise to two basic consequences:
(i) that Euclidean manifolds which have the same translations in 
their covering groups exhibit the same spike spectrum of topological
nature;  
(ii) that individual pair separation histograms (PSH) of hyperbolic 
3-manifolds exhibit no spike of topological origin, since there are
no Clif\/ford translations in the hyperbolic geometry.
{}From the expression for the EPSH they have 
found it is also apparent that the contribution 
due to the multiply-connectedness (topological signature) 
must arise in PSH's even when there are only a few 
images for each object.%
\footnote{Note however, that whether or not
the signatures can be distinguished from the statistical noise
when studying a \emph{single} observational catalogue is an
independent question.} 
These results have been formally derived 
{}from very general f\/irst principles in~\cite{GTRB98}, nevertheless
they have not been thoroughly clarif\/ied through concrete simulations.

In the quest for a means of reducing the statistical f\/luctuations
well enough to clear up the signal of non-translational isometries
in PSH's, Gomero \emph{et al.\/}~\cite{GTRB98} also suggested the
mean pair separation histogram (MPSH) as a very f\/irst approach
to ref\/ine upon the crystallographic method. But, although 
the MPSH procedure rests on a well known result from elementary 
statistics, apart from the cases recently discussed in%
~\cite{GRT99,GRT00}, no other MPSH has been explicitly built 
to show that the MPSH scheme is useful in simulations, 
and that it is a suitable approach for studying the role of 
non-translational isometries as well as the topological signature
of any 3-manifold of constant curvature.

The most patent evidence of multiply-connectedness in PSH's is the 
presence of topological spikes, which arise  when the isometry 
is a Clif\/ford translation. The other  isometries, however, 
manifest  as small deformations of the EPSH corresponding to the 
underlying simply connected manifold. In computer-aided simulations, 
however,  histograms contain statistical noise, which on the 
one hand can give rise to sharp peaks of purely statistical 
origin~\cite{FagundesGausmann98,GTRB98}, on the other hand it 
can mask or even hide the tiny deformations due to non-translational 
isometries. 
The most immediate approach to cope with these problems clearly 
is through the reduction of the noise in PSH's by using, for 
example, the MPSH scheme. Another possible way of facing them 
is by focusing in a more appropriate quantity rather than PSH's. 
Obviously one can also combine these approaches to tackle 
those problems.

In this article, we f\/irstly recast the theoretical 
results obtained in~\cite{GTRB98} to show how one can extract the 
topological signature of any multiply connected manifold of constant 
curvature by using a suitable quantity, which turns out to be a constant 
factor times the dif\/ference $\Phi_{exp}(s_i) - \Phi^{sc}_{exp}(s_i)$ of 
the EPSH corresponding to the multiply connected manifold minus the
EPSH of the underlying simply connected covering manifold.
Secondly we show, through concrete simulations and based upon our 
theoretical results, two ways of obtaining (extracting and constructing)
the topological signatures of multiply connected manifolds of constant
curvature. There also emerges from our theoretical results and simulations 
how separated contributions of the covering isometries are composed to 
give rise to the (complete) topological signature of a specif\/ic 
manifold.
Further, we make clear that the topological signature arises 
(in the MPSH, i.e. in the mean)
even when there are just a few images for each object.
Finally, it is also illustrated that the MPSH procedure works by 
building MPSH's from simulated catalogues, and that it is a 
suitable approach for studying the topological signature as
well as the role of non-translational isometries.

The scope of this paper is as follows. In the next section we
set the notation, brief\/ly  recast the major result of 
ref.~\cite{GTRB98}, and presents the theoretical grounds of
the two approaches for obtaining (extracting and constructing) 
the topological signatures of any multiply connected 3-manifolds 
of constant curvature. In section~\ref{simula} we present and 
discuss simulations that make apparent the following issues:
(i) that the two procedures for obtaining the topological signature 
of multiply connected 3-manifolds discussed in section~\ref{sig}
work, and give rise to the same signature;
(ii) how the separated contributions of the covering isometries 
add up to form the plain topological signature of RW universes; and
(iii) that the topological signature arises in simulations 
(in the mean, i.e. through MPSH's)
even when there are just a few images for each object.
It is also shown in that section that the mean pair separation 
histogram technique works and that it is a suitable approach 
for studying the topological signature as well as the role of 
non-translational covering isometry of any multiply connected 
3-manifold of constant curvature (spacelike sections of all RW 
universes). In section~\ref{remarks} we summarize our main
results and brief\/ly indicate possible approaches for further 
investigations.

\section{Topological signature}
\label{sig}
\setcounter{equation}{0}

In this section we shall f\/irst set the notation and brief\/ly  
recast the major result of ref.~\cite{GTRB98} so as to show how 
one can extract a topological signature of any multiply connected 
manifold of constant curvature. 

Let us start by recalling that a catalogue $\mathcal{C}$ is a 
set of \emph{observed} images, subset of the \emph{observable} 
images, which are clearly contained the observable universe,
which in turn is the part of the universal covering manifold 
$\widetilde{M}$ causally connected to an image of a given 
observer.

Consider a catalogue $\mathcal{C}$ with $n$ cosmic images and 
denote by $\eta(s)$ the number of pairs of images whose  
separation is $s$. Consider also that our observed universe is a 
ball of radius $a$ and divide the interval $(0,2a]$ in $m$
equal subintervals $J_i$ of length $\delta s = 2a/m$. Each of
such subintervals has the form
\begin{equation} \label{Ji}
J_i = (s_i - \frac{\delta s}{2} \, , \, s_i + \frac{\delta
s}{2}] \qquad ; \qquad i=1,2, \,\ldots\, ,m \;\, ,
\end{equation}
and is centred at
\begin{displaymath}
s_i = \,(i - \frac{1}{2}) \,\, \delta s \;.
\end{displaymath}
The PSH is a normalized function which counts the number of pair 
of images separated by a distance that lies in the 
subinterval $J_i$. Thus the function PSH is given by
\begin{equation}  \label{defpsh}
\Phi(s_i)=\frac{2}{n(n-1)}\,\,\frac{1}{\delta s}\,
               \sum_{s \in J_i} \eta(s) \equiv 
               \sum_{s \in J_i} \widehat{\eta}\,(s) \;,
\end{equation}
and  is clearly subjected to the normalizing condition
\begin{equation}
\sum_{i=1}^m \Phi(s_i)\,\, \delta s = 1 \; .
\end{equation}

In a multiply connected universe the periodic distribution of images 
on $\widetilde{M}\,$ due to the covering group gives rise to 
correlations in their positions, and these correlations can be 
couched in terms of correlations in distances between the pairs 
of images. The examination of the behaviour of these distance 
correlations can be made as follows.

If one considers an ensemble of comparable catalogues 
with the same number $n$ of images, and corresponding to the same 
3-manifold $M$ of constant curvature, one can compute probabilities 
and expected values of quantities which depend on the images in 
the catalogues of the ensemble. Clearly each catalogue of the 
ensemble gives a set of images distributed in the observed universe.
In particular, one can compute the most important quantity for our
purpose here, which is the expected number, $\eta_{exp}(s_i)$, 
of pairs of cosmic images in a catalogue $\mathcal{C}$ of the 
ensemble with separations in $J_i$. 
This quantity is quite relevant because from it one has the 
expected (normalized) pair separation histogram which clearly is 
given by
\begin{equation}
\label{defepsh}
\Phi_{exp}(s_i) = \frac{2}{n(n-1)}\,\,\frac{1}{\delta s} \,\,
\eta_{exp}(s_i) = \frac{1}{N}\,\,\frac{1}{\delta s}\,\,
\eta_{exp}(s_i) \; ,
\end{equation}
where obviously $N=n(n-1)/2$ is the total number of pairs of cosmic 
images  in $\mathcal{C}$, and so  the  whole coef\/f\/icient of 
$\eta_{exp}(s_i)$ is nothing but a normalizing factor.

If one denotes by $F_g(s_i)$ and $F_u(s_i)$, respectively, the 
probability that the images of a $g$-pair%
\footnote{In line with the notation of~\cite{GTRB98}, when referring 
collectively to correlated pairs we use 
the terminology $\Gamma$-pairs, reserving the name $g$-pair for a 
correlated pair corresponding to a specif\/ic isometry $g \in \Gamma$.
In other words, a $g$-pair is a pair of the form $(p, gp)$ for any 
(f\/ixed) isometry $g$. It should also be noted that the $\Gamma$-pairs
are the same as the \emph{type II pairs} of Uzan 
\emph{et al.}~\cite{UzLeLu99}.} 
and an uncorrelated 
pair be separated by a distance that lies in $J_i$, the expected 
number $\eta_{exp}(s_i)$ can be expressed as
\begin{equation} \label{nexp1}
\eta_{exp}(s_i) = N_u \, F_u(s_i) + \frac{1}{2}\,
\sum_{g \in \widetilde{\Gamma}} \,N_g \, F_g(s_i) \;,
\end{equation} 
where  $\widetilde{\Gamma}$ denotes the covering group 
$\Gamma$ without the identity map, and where $N_u$ and $N_g$
denote, respectively, the (total) expected  number of uncorrelated 
pairs and the (total) expected  number of $g$-pairs in a typical 
catalogue $\mathcal{C}$ of the ensemble.%
\footnote{For a formal proof that the decomposition~(\ref{nexp1}) 
can always be made see~\cite{GTRB98}.}

Inserting eq.~(\ref{nexp1}) into eq.~(\ref{defepsh}) 
one obtains
\begin{equation} \label{EPSH1}
\Phi_{exp}(s_i) = \frac{N_u}{N}\,\,\frac{1}{\delta s}\,\,F_u(s_i)
      + \frac{1}{2}\,\sum_{g \in \widetilde{\Gamma}} 
         \,\, \frac{N_g}{N}\,\, \frac{1}{\delta s}\,\, F_g(s_i) \;.
\end{equation}
However, from equation~(\ref{defepsh}) one is led to def\/ine 
the EPSH's corresponding to uncorrelated pairs and associated 
to an isometry $g$, respectively, as 
\begin{eqnarray} 
\Phi^{u}_{exp}(s_i) &=& \frac{1}{N_u}\,\frac{1}{\delta s}\,\, 
        \eta^u_{exp}(s_i) = \frac{1}{\delta s}\, F_u(s_i) 
                                               \label{epshu}\;,\\
\Phi^{g}_{exp}(s_i) &=& \frac{1}{N_g}\,\frac{1}{\delta s}\, \,
          \eta^g_{exp}(s_i)  =\frac{1}{\delta s} \,F_g(s_i) 
                                                 \label{epshg} \;, 
\end{eqnarray}
where clearly $N_u = \sum_{s_i} \eta^u_{exp}(s_i)\,$ and 
$\,N_g = \sum_{s_i} \eta^g_{exp}(s_i)\,$.

Thus using (\ref{epshu}) and (\ref{epshg}), equation (\ref{EPSH1}) 
reduces to either of the following forms:
\begin{equation}   \label{EPSH2}
\Phi_{exp}(s_i) = \mu_u \,\, \Phi^{u}_{exp} (s_i) 
+ \sum_{g \in \widetilde{\Gamma}}\,  \mu_g \,\, \Phi^g_{exp}(s_i) 
\quad \mbox{with} \quad 
\mu_u=\frac{N_u}{N}\,, \quad  \mu_g=\frac{1}{2}\,\frac{N_g}{N} \;, 
\end{equation}
or equivalently
\begin{equation} \label{EPSH3}
\Phi_{exp}(s_i)  =  \frac{1}{n-1}\,\, [\,\nu_u\,\Phi_{exp}^{u}(s_i) + 
\,\sum_{g \in \widetilde{\Gamma}}\, \nu_g\, \Phi_{exp}^g(s_i) \,] 
\quad \mbox{with} \quad  
\nu_u = 2\,\frac{N_u}{n}\,, \quad \nu_g =\frac{N_g}{n} \;. 
\end{equation}

For simply connected manifolds $M$, since all $N$ pairs are 
uncorrelated, equation~(\ref{defepsh}) reduces to 
\begin{equation} \label{epshsc}
\Phi^{sc}_{exp}(s_i) = \frac{1}{N}\,\frac{1}{\delta s}\,\,
                            \eta^{sc}_{exp}(s_i)
                       = \frac{1}{\delta s} \,F_{sc}(s_i) \;,
\end{equation}
where $F_{sc}(s_i)$ is the probability that the two objects 
in $M$ be separated by a distance that lies 
in $J_i$.%
\footnote{It should be noticed that although all pairs in the 
catalogues corresponding to simply connected manifolds are 
uncorrelated the probabilities $F_u(s_i)$ and $F_{sc}(s_i)$ 
are not the same.}

Now since the pairs of cosmic images are either correlated 
($\Gamma$-pairs) or uncorrelated we must have
\begin{equation} \label{sumNuNg}
N_u + \frac{1}{2}\,\sum_{g \in \widetilde{\Gamma}}\,N_g\,= \,N\;,
\end{equation}
which on the one hand leads to  
\begin{equation}    \label{sumug}
\mu_u + \sum_{g \in \widetilde{\Gamma}}\,\mu_g = 1 \;,
\end{equation}
and on the other hand it gives rise to an alternative expression for 
the EPSH $\Phi_{exp}(s_i)\,$ in terms of $\Phi^{sc}_{exp}(s_i)\,$. 
Indeed, using~(\ref{EPSH2}) and (\ref{sumug}) one easily 
obtains
\begin{equation}   \label{EPSH4}
\Phi_{exp}(s_i) = \Phi^{sc}_{exp}(s_i)
  + \mu_u\,[\,\Phi^{u}_{exp}(s_i) - \Phi^{sc}_{exp}(s_i)\,] 
  + \sum_{g \in \widetilde{\Gamma}}\, 
\mu_g \, [\,\Phi^g_{exp}(s_i) - \Phi^{sc}_{exp}(s_i)\,] \;.
\end{equation}

\begin{sloppypar} 
{}From equation~(\ref{EPSH4}) one can obtain an expression for  
the \emph{topological signature} $\varphi^S(s_i) \equiv (n-1)\,
[\Phi_{exp}(s_i)  - \Phi^{sc}_{exp}(s_i)]\,$. Indeed, 
using~(\ref{EPSH4}) together with the def\/inition of 
$\mu_u$ and $\mu_g$ one easily obtains
\end{sloppypar} 
\begin{equation} \label{topsig1}
\varphi^S(s_i) = 
(n-1)\,[\,\Phi_{exp}(s_i) - \Phi^{sc}_{exp}(s_i)\,] = 
\varphi^U(s_i) + \varphi^{\Gamma}(s_i) \;,
\end{equation}
where 
\begin{equation}   \label{PhiU}
\varphi^U(s_i) = \nu_u \left[\,\Phi^{u}_{exp}(s_i) 
             - \Phi^{sc}_{exp}(s_i)\,\right]  \;,
\end{equation}
and
\begin{equation}   \label{PhiGamma}
\varphi^{\Gamma}(s_i) = \,\sum_{g \in \widetilde{\Gamma}} 
\nu_g\,[\, \Phi^g_{exp}(s_i) - \Phi^{sc}_{exp} (s_i)\,] \;, 
\end{equation}
and where we have used the def\/initions of
$\nu_u$ and $\nu_g$ given by (\ref{EPSH3}).

An important word of clarif\/ication is in order here: 
although throughout this article we loosely use the terminology 
topological signature of a manifold and of a universe, it should 
be stressed that the topological signature $\varphi^S(s_i)$ depends 
on the pair 3-manifold and observed universe $\mathcal{B}_a\,$, as 
well as on the relative position of the FP with respect to 
$\mathcal{B}_a\,$.

To show how one can extract through computer-aided simulations
the topological signature $\varphi^S(s_i) \equiv (n-1)\,
[\Phi_{exp}(s_i)  - \Phi^{sc}_{exp}(s_i)]\,$ of multiply 
connected universes, a quite important point to be noted is that 
the EPSH is essentially a typical PSH from which the statistical 
noise has been withdrawn. Hence we have
\begin{equation}  \label{noise1}
\Phi(s_i) = \Phi_{exp}(s_i) + \rho(s_i) \; ,
\end{equation}
where $\Phi(s_i)$ is a typical PSH constructed from $\mathcal{C}$
and $\rho(s_i)$ represents the statistical f\/luctuation that
arises in the PSH $\Phi(s_i)$. Alternatively from~(\ref{noise1})
one has that a PSH is an EPSH plus the statistical f\/luctuation.

Using now the decomposition~(\ref{noise1}) together 
with~(\ref{topsig1}) one easily obtains
\begin{equation}   \label{nonpropquant} 
(n-1)\,\,[\,\Phi(s_i) - \Phi^{sc}(s_i)\,] = 
    \varphi^{S}(s_i) - \,(n-1)\,[\,\rho(s_i)-\rho^{sc}(s_i)\,] \;,
\end{equation}
where $\rho(s_i)\,$  and $\rho^{sc}(s_i)$ are the statistical
noises which arise in the PSH's corresponding to the multiply connected 
and the covering manifolds, respectively.

Clearly the right-hand side of (\ref{nonpropquant}) gives the topological 
signature $\varphi^{S}(s_i) $ intermixed with the two statistical 
f\/luctuations $\rho(s_i)\,$  and $\rho^{sc}(s_i)$.
In practice, however, one can obviously approach to the suitable 
quantity $[\Phi_{exp}(s_i)  - \Phi^{sc}_{exp}(s_i)]\,$  
by reducing the statistical f\/luctuations,
through any suitable statistical method to lower the noises.  
The simplest way to accomplish this is to use several comparable 
catalogues, with approximately equal number of cosmic images, 
for the construction of a \emph{mean} pair separation histogram 
(MPSH). For suppose we have $K$ computer-generated catalogues 
$\mathcal{C}_k$ ($k=1,2,\dots,K$) whose PSH's for a given value
of $m$ are given by
\begin{equation}  \label{sample-PSH}
\Phi_k(s_i) = \frac{2}{n_k(n_k-1)}\, \frac{1}{\delta s}\,
\sum_{s \in J_i} \eta_k(s) \;,
\end{equation}
where $n_k$ is the number of images in the catalogue 
$\mathcal{C}_k$. The MPSH def\/ined by
\begin{equation} \label{meanPhi}
<\!\Phi(s_i)\!>\; = \frac{1}{K} \,\sum_{k=1}^K \Phi_k(s_i)
\end{equation}
is such that in the limit $K \to \infty\,$ one has 
$\Phi_{exp}(s_i) \simeq \, <\!\Phi(s_i)\!>\,$. 
Elementary statistics tells us that the statistical 
f\/luctuations in the MPSH are reduced by a factor of 
$1/\sqrt{K}$, which makes clear that MPSH is a suitable 
approach to deal with the above-mentioned f\/luctuations.

In brief, the use of the MPSH to extract the topological
signature $\varphi^S(s_i)$ consists in the use of $K$ (say) 
computer-generated comparable  catalogues (with approximately
the same number $n$ of images  and corresponding to the
same manifold $M$) to obtain the mean pair separation histogram
$<\!\Phi(s_i)\!>\,$ and analogously to have $<\!\Phi^{sc}(s_i)\!>\,$; 
and use them as an approximation for $\Phi_{exp}(s_i)$ and 
$\Phi^{sc}_{exp}(s_i)\,$, to construct the topological signature 
$\varphi^S(s_i) \simeq
(n-1)\,[<\!\Phi(s_i)\!>  - <\!\Phi^{sc}(s_i)\!>]\,$.
Clearly the greater is the number $K$ of catalogues the lower 
are the statistical noises $\rho(s_i)\,$, $\rho^{sc}(s_i)\,$  
and obviously the better are the approximations 
$<\!\Phi(s_i)\!> \,\simeq  \Phi_{exp}(s_i)\,$ and 
$<\!\Phi^{sc}(s_i)\!> \,\simeq \Phi^{sc}_{exp}(s_i)\,$.

An improvement of the above procedure to extract the topological 
signature $\varphi^S(s_i)$ comes out for the cases one can 
derive the expression for the PSH's $\,\Phi^{sc}_{exp}(s_i)$
corresponding to the simply connected covering manifolds. 
In a recent work~\cite{BT99} (see also~\cite{Reboucas00}) the explicit 
formulae for the expected pair separation histogram functions 
$\Phi^{sc}_{exp}(s_i)$ corresponding to an uniform distribution 
of objects in the covering manifolds $E^3$, $H^3$ and $S^3$ have 
been obtained.%
\footnote{Actually, they have found the explicit expressions for 
the EPSH corresponding to each (Euclidean, hyperbolic and elliptic)
observed (homogeneous) universe, which 
is a spherical ball $\mathcal{B}_a$ with radius $a$ fulf\/illed with 
an uniform distribution of cosmic objects.} 
Thus, for such universes with homogeneous distribution 
of objects one has from the very beginning $\rho^{sc}(s_i)=0\,$, 
and obviously the topological signature can be rewritten as
$\varphi^S(s_i) \simeq (n-1)\,[\,<\!\Phi(s_i)\!> - \,\,
\Phi^{sc}_{exp}(s_i)\,]\,$. 

For the sake of selftcontainedness we present below for use in
section~\ref{remarks} the explicit expressions for the EPSH's found 
in refs.~\cite{BT99,Reboucas00} corresponding to simply connected Euclidean, 
hyperbolic and elliptic universes with a uniform distribution 
of cosmic objects. 
 
\vspace{4mm}
\noindent \textbf{Euclidean Universes}
\begin{equation}  \label{fdensE}
\Phi^{sc}_{exp}(a,s)=\frac{3}{16\,a^6}\,\,s^2\,(2a-s)^2\,(s+4a)\;,
\end{equation} 
which holds for $s\in (0, 2a]\,$.

\vspace{3mm}
\noindent \textbf{Hyperbolic Universes}
\vspace{1mm}
\begin{equation}  \label{fdensH}
\Phi^{sc}_{exp}(a,s)=\frac{8 \sinh^2 s}{(\sinh 2a -2a)^2}\,\, \left[
\,\cosh a \,\,\mbox{sech}(s/2)\,\sinh(a-s/2)\,\,-\,(a-s/2)\,\right]\;, 
\end{equation}  
which also holds for $s\in (0,2a]$.

\vspace{3mm}
\noindent \textbf{Elliptic Universes}
\begin{eqnarray} 
\Phi^{sc}_{exp}(a,s) & = & \frac{8\sin^2 s}{(2a-\sin 2a)^2} \,\,\left\{
   2a-\sin 2a-\pi + \Theta(2\pi-2a-s)\,\,\left[\,\sin 2a\,+\,\pi
            \right. \right. \nonumber \\ 
& & \qquad \qquad \qquad \quad
      -\,a-s/2 - \cos a\,\, \sec(s/2)\,\,\sin(a-s/2)\, \left. \right]\, 
                     \left.\right\} \label{fdensEL} \;,
\end{eqnarray}
which holds for all $a\in (0,\pi]$ and $s\in (0,\mbox{min}(2a,\pi)]\,$,
and where $\Theta$ denotes the well-known Heaviside function.  

It should be noticed that in~(\ref{fdensEL}) as well as in (\ref{fdensE}) 
and (\ref{fdensH}) we have made explicit that the EPSH's depend upon 
the radius $a$ of the observed universe $\mathcal{B}_a\,$.

{}From equations~(\ref{topsig1})~--~(\ref{PhiGamma}) it is also clear 
that another approach to obtain (constructive approach) the topological 
signature of a multiply connected manifold is by considering the sum 
of the terms on the right-hand side of~({\ref{topsig1}), namely
$\varphi^U(s_i)=\nu_u\, [\Phi^{u}_{exp}(s_i) - \Phi^{sc}_{exp}(s_i)]$ 
and $\varphi^{\Gamma}(s_i) = \sum_{g \in \widetilde{\Gamma}} 
\nu_g\,[\, \Phi^g_{exp}(s_i)- \Phi^{sc}_{exp} (s_i)\,]$.
In the next section we shall perform simulations to obtain (construct) 
the topological signature of a given manifold using also this 
alternative procedure and make clear that one ends up with the
same signature obtainable through the f\/irst approach,
as it should be expected from the outset. In doing so we 
will be showing in addition that indeed~(\ref{topsig1}),
with (\ref{PhiU}) and (\ref{PhiGamma}), holds within the
accuracy of our plot, of course.  

To close this section it should be noted that in a recent work, 
in studying PSH's corresponding to one of the Best manifold, 
Fagundes and Gausmann~\cite{FagGaus} have considered the similar 
quantity $\Phi(s_i) - \Phi^{sc}(s_i)$, instead of the quantity 
$\Phi_{exp}(s_i) - \Phi^{sc}_{exp}(s_i)$ we are interested
here. However, apart from the fact that they have not presented any 
theoretical argument sustaining their suggestion, from
(\ref{nonpropquant}) one has that their scheme gives rise to 
a fraction $1/(n-1)$ of the topological signature $\varphi^{S}(s_i) $ 
plus (algebraically) the f\/luctuations corresponding to both PSH's 
involved, namely that which arises from the multiply connected Best 
manifold and that originated from the covering space. 

\begin{sloppypar}
\section{Topological signature from computer-aided simulations}
\label{simula}
\setcounter{equation}{0}
\end{sloppypar}

In this section we shall present and analyze simulations that 
clarify the following points: 
(i) how one can obtain (extract and/or construct) the topological 
signature of multiply connected manifolds; 
(ii) how separated terms which arise from the covering isometries 
are composed to form the topological signature of a manifold; and
(iii) show that the topological signature arises in simulations 
(in the mean)
even when there are just a few images for each object.
In the clarif\/ication of theses points we shall use the MPSH 
technique to make apparent that it is a suitable approach to study 
the topological signature and the role of non-translational 
isometries as it has been stated in~\cite{GTRB98} and illustrated
in~\cite{GRT99} for the case of f\/lat manifolds.

The f\/irst series of computer-aided simulations concerns a compact 
orientable Euclidean manifold of class ${\mathcal G}_6$ in Wolf's 
classif\/ication~\cite{Wolf}. Recently it has been found a set 
of expressions for the face-pair generators of a fundamental 
polyhedron (FP) of a cubic manifold of this class. 
We shall denote this Euclidean manifold by ${\mathcal T}_4$ 
in agreement with the notation used in~\cite{GRTB99} and \cite{BGRT98}, 
wherein the cubic FP and the pairwise faces identif\/ication are shown 
in f\/igure~1. 
Relative to a coordinate system whose origin coincides with the 
center of the FP, the actions of the generators $\alpha$, $\beta$ 
and $\delta$ on a generic point $p = (x, y, z)$ were shown to be 
described by~\cite{Wolf,Gomero}
\begin{eqnarray}  
\alpha \, p  & = &  (x+L, \,-y, \,-z) \;, \label{isoa}   \\
\beta \, p  & = &  (-x,\, z+L,\, y)   \;, \label{isob} \\
\delta \, p  & = &  (-x,\, z, \, y+L) \; , \label{isod} 
\end{eqnarray}
where $L$ is the edge of the cubic FP. Clearly the actions of the
inverses of these generators are given by
\begin{eqnarray}   \label{gisoinv}
\alpha^{-1} \, p  & = &  (x-L, \,-y, \,-z) \;, \label{isoa-1}  \\
\beta^{-1}\, p  & = &  (-x,\, z,\, y-L)   \;, \label{isob-1} \\
\delta ^{-1}\, p  & = &  (-x,\, z-L, \, y) \;. \label{isod-1} 
\end{eqnarray}

In the simulations corresponding to the cubic manifold 
${\mathcal T}_4$ the centre of the FP was taken to be the origin
of the coordinate system, and to coincide with the centre of the 
observed universe $\mathcal{B}_a\,$, whose diameter $2a$ is
such that  $2 a = L \, \sqrt{2} \simeq 1.41 \,L \,$.
It should be noted that with this ratio for $a/L$ and for
$s \in (0,2a)\,$ one has only the contribution of non-translational 
isometries for the topological signature. Indeed, according to  
eqs.~(\ref{isoa})~--~(\ref{isod-1}) the translations of shortest 
length are due to $\beta^2$, $\delta^2$ and their inverses, and 
are translations of $L\,\sqrt{2}$ units. So, as we shall take  
$0 < s < 2a$  no spike will appear in our PSH's for this cubic 
manifold.%
\footnote{Note, en passant, that the topological signature depends
on the pair: 3-manifold and observed universe $\mathcal{B}_a\,$,
and also on the relative position of the FP with respect to 
observed universe $\mathcal{B}_a\,$.
So, had we considered, for example, a ball with double diameter  
$4a=2\,\sqrt{2} \simeq 2.83 $, and kept the same relative 
position of FP and $\mathcal{B}_a\,$, the topological signature 
$\varphi^S(s_i)$ would exhibit spikes at $s=\sqrt{2} \simeq 1.41 $, 
$s=2$, and $s=\sqrt{6} \simeq 2.45\,$ and so forth 
(see ref.~\cite{GRT99} for more details).}  

To show how one can extract the topological signature 
$(n-1)\,[\Phi_{exp}(s_i) - \Phi^{sc}_{exp}(s_i)]= 
\varphi^{S}(s_i)\,$ of $\,{\mathcal T}_4\,$, as well as
to illustrate the MPSH procedure, and to make clear that the 
topological signature arises in simulations when there are a few 
images of some cosmic objects, we have written a program whose input 
are the number $K$ of catalogues, the radius $a$ of the observed 
universe $\mathcal{B}_a\,$, the number $m$ of subinterval (bins), 
and the number $n_s$ of objects inside the FP (seeds).
The program generates $K$ \emph{different} catalogues,  starting 
(each) from the same number $n_s$ of homogeneously distributed
seeds inside the FP. For each  bin $J_i$ of width $\delta s = 2a/m$ 
it counts the normalized number of pairs $\sum \widehat{\eta}_k\,(s)$ 
for all catalogues $k$ from 1 to $K$. 
Finally, it calculates the normalized average numbers of pairs
for all $s_i \in (0,2a)\,$, f\/inding therefore, according to~%
(\ref{meanPhi}), the mean pair separations histogram 
$\,<\!\Phi(s_i)\!>\,$ over $K$ catalogues. 

Now, on the one hand we know that the mean pair separation histogram 
$<\!\Phi(s_i)\!>\,$ is approximately equal to the expected pair separation 
histogram $\Phi_{exp}(s_i)$ for large enough number $K$. On the other 
hand, we have that the explicit expression for $\Phi^{sc}_{exp}(s_i)$ 
corresponding to a uniform distribution of objects in a Euclidean
universe is given by eq.~(\ref{fdensE}). Thus, the plot of 
$(n-1)\,[\,<\!\Phi(s_i)\!> - \,\,\Phi^{sc}_{exp}(s_i)\,]\,$
gives, up to a statistical f\/luctuation $\rho(s_i)$, a  
topological signature of the ${\mathcal T}_4$. 
As a matter of fact, since not all catalogues generated from a 
f\/ixed number of seeds inside the FP have the same number $n$ of 
images, in our program instead of $n$ we have used the average 
number $\,<\!n\!>\,$ of images per catalogue to plot 
the topological signature $\varphi^S(s_i)\,$.

\begin{sloppypar}
Using the above-described program we have performed simulations 
for the manifold ${\mathcal T}_4$ with $L=1$ in an observed 
universe with radius $a = \sqrt{2}/2  \simeq 0.71 \,$, 
$\delta s =0.01$, and with dif\/ferent number $n_s$ of
seed objects uniformly distributed in the FP.
Figure~1a is the graph of the topological signature
$(<\!n\!> -\,1\,)\,[<\!\Phi(s_i)\!> - \,\Phi^{sc}_{exp}(s_i)]$ 
for the manifold ${\mathcal T}_4$ 
and was obtained using the MPSH procedure for $K=16000$ 
catalogues, and  $n_s=15$, which corresponds to an average 
number of images per catalogue $<\!n\!>\,\simeq 23$. 
Figure~1b shows a graph of the topological signature for the 
same manifold and obtained through the MPSH scheme for 
identical number of catalogues, but now the number of seeds was 
taken to be $n_s=100$,  which corresponds to $<\!n\!>\, \simeq 153$ 
images. 
These f\/igures clarify the following relevant points:
First, that the MPSH approach is indeed a suitable approach 
to reveal the topological signatures of non-translational 
isometries; Second, as long as one can extract the topological
signature for a small numbers of seeds ($n_s=15$) and it 
is essentially the same obtained for a fairly large 
number of seeds ($n_s=100$), it becomes clear from these
concrete simulations that the plain topological signature 
arises in simulations where there are just a few images 
for each object. 
\end{sloppypar}

According to our earlier discussion, from equation~%
(\ref{topsig1})~--~(\ref{PhiGamma}) it also is clear 
that an alternative approach to obtain (constructive approach)
the topological signature of a multiply-connected manifold is 
to consider the sum of the terms on the right-hand side 
of~({\ref{topsig1}), namely $\varphi^U(s_i) =
\nu_u\, [\Phi^{u}_{exp}(s_i) - \Phi^{sc}_{exp}(s_i)]$ 
and $\varphi^{\Gamma}(s_i) = \sum_{g \in \widetilde{\Gamma}} 
\nu_g\,[\, \Phi^g_{exp}(s_i)- \Phi^{sc}_{exp} (s_i)\,]$.
We have also performed simulations following this approach to 
determine the topological signature of  the manifold ${\mathcal T}_4$ 
again with $L=1$ and $a = \sqrt{2}/2  \simeq 0.71 \,$. 
In doing so we intend, on the one hand, to numerically verify 
equation~(\ref{topsig1}), on the other hand, to illustrate how the 
separated terms which arise from the covering isometries sum up 
to give the topological signature.
In what follows we shall report more details of the simulations
involved in this second approach.

To make explicit the contribution $\Phi^{g}_{exp}(s_i)$ of a specif\/ic
isometry $g = \alpha$ (say) to the topological signature term 
$\varphi^{S}(s_i)\,$ [see eqs.~(\ref{epshg}) and 
(\ref{topsig1})~--~(\ref{PhiGamma})] in computer-generated 
catalogues we have written a program whose inputs are 
the total number $N_g$ of correlated pairs, the length $L$,
the radius $a$ of the universe $\mathcal{B}_a\,$ and the number
$m$ of subintervals (bins); and starts by setting a counting 
variable $k=0$ and has the following steps:
\begin{enumerate} 
\item[(i)]  
randomly and homogeneously take an object $p$ (say) inside FP 
\item[(ii)]  
apply the isometry $\alpha$ and check if  
the image $\alpha p$ is still inside the ball $\mathcal{B}_a\,$. 
If no, return to step (i), if yes let $k=k+1$ and go to 
step 
\item[(iii)]  
calculate the distance $s$ between $p$ and 
$\alpha p$, then 
\item[(iv)]  
call a procedure that computes for all 
$s \in (0, 2a)$ the normalized sum of pairs $\sum \widehat{\eta}(s)$
whose images are separated by a distance that lies in a bin $J_i$ 
with width $\delta s = 2a/m$;
\item[(v)] 
check if the $k$ is equal to $N_g$. If no, return to step (i); 
otherwise plot the sequence of pairs 
$[\,x = (j-1/2)\,\delta s\,, \,\, y = \sum \widehat{\eta}(s_i) /N_g\,]$ 
for $\,j=1, \ldots ,m$  and stop.
\end{enumerate} 
We have used this program to compute the contribution of the
isometries $\alpha$ and $\beta$ for $N_g=16000$, $m=142\,$ 
($\delta s =0.01$), $\,a=\sqrt{2}/2$  and $L=1$. Figures~2a and~2b 
show, respectively, the graphs of $<\!\Phi^{\alpha}(s_i)\!>$ and 
$<\!\Phi^\beta(s_i)\!>\,$ for the isometries $\alpha$ and $\beta$,
whose actions are given, respectively, by (\ref{isoa}) 
and~(\ref{isob}). The fact that the action of isometry
$\alpha$, for example, gives rise to $g$-pairs whose minimum 
separation is $1$ can be easily understood from the expression 
for the distance $d$ between $p$ and $\alpha\,p$, namely from 
$d(p,\alpha p)\,$. Indeed it is trivial to show that 
$d(p,\alpha p) \geq 1$. Similarly, form~(\ref{isob}) one clearly
has $d(p,\beta p) \geq \sqrt{2}/2 \simeq 0.71\, $, making clear
that the $\beta$-pairs have also a minimum separation.%
\footnote{Actually, it can be explicitly shown that the 
set of starting points at which the  non-translational
isometries begin to contribute for the topological 
signature are given by $d^2 = 0.5, 1, 4.5, 9, 12.5$, 
and so forth~\cite{GRT99}. Thus, it is clear that the inclusion 
or exclusion of the contribution of a specif\/ic isometry 
depends on the radius $a$ of the observed universe 
$\mathcal{B}_a\,$, reinforcing the fact that the topological
signature depends upon the ball $\mathcal{B}_a\,$.}

For the ratio $a/L= \sqrt{2}/2$ and relative position of FP and 
$\mathcal{B}_a\,$ that we have chosen, only the isometries given 
by~(\ref{isoa})~--~(\ref{isod-1}) will contribute to the term 
$\varphi^\Gamma(s_i)$. Furthermore, from these equations one clearly 
has $\Phi^{\alpha^{-1}}_{exp}(s_i)=\Phi^\alpha_{exp}(s_i)$ 
and $\Phi^{\beta^{-1}}_{exp}(s_i)= \Phi^\beta_{exp}(s_i)$. Again,
due to the symmetry of eqs.~(\ref{isoa})~--~(\ref{isod-1}) one easily
obtains that $\Phi^\delta_{exp}(s_i) = \Phi^{\delta^{-1}}_{exp}(s_i)= 
\Phi^\beta_{exp}(s_i)\,$. Therefore, the second term on the 
right hand side of~(\ref{topsig1}) reduces to  
\begin{equation}  \label{PhGama}
\varphi^{\Gamma}(s_i)\simeq \nu_g\,[\,2 <\!\Phi^{\alpha}(s_i)\!>
+ \, 4 <\!\Phi^\beta(s_i)\!> - \,6\,\Phi^{sc}_{exp}(s_i)\,] \;.
\end{equation}
Finally, for $a/L$ and the relative position of  FP and 
$\mathcal{B}_a\,$ we have chosen, according to equations~%
(\ref{isoa})~--~(\ref{isod-1}) one easily has that for an uniform 
distribution of objects the coef\/f\/icients $\nu_g$ are equal 
for all isometries $g=\alpha, \beta, \delta$ and their inverses. 
The computation of the value $\nu_g$ reduces to the calculation 
of quotient of volumes, and turns out to be $\nu_g  \simeq 0.1161165234\,$ 
(for more details on how one can calculate $\nu_g$ see~\cite{GRT99}).
Figure~3 shows the graph of $\Phi^{sc}_{exp}(s_i)$ for $a=\sqrt{2}/2$ 
as given by~(\ref{fdensE}). 
Figure~4a shows the graphs of this $\varphi^{\Gamma}(s_i)$, given
by~(\ref{PhGama}), for $\delta=0.01$, $a=\sqrt{2}/2$ and $L=1$. 
It should be stressed that this f\/igure is nothing but 
the above-described suitable combination of f\/igures~2 and~3
multiplied by the constant factor $\nu_g\,$.

Again for the same relative position between FP and $\mathcal{B}_a\,$, 
the contribution of the term $\varphi^U (s_i)$ to the topological 
signature [$\,\varphi^S(s_i)\,$] of ${\mathcal T}_4$ in shown in 
f\/igure~4b, and was obtained for the following set of inputs: 
$m=142$, $a=\sqrt{2}/2$, $L=1$, and $K=16000$.

Figure~5 gives the sum of the contributions due to
$\varphi^{\Gamma}(s_i)$ and $\varphi^U(s_i)\,$. The comparison 
between f\/igures~1a and~1b with f\/igure~5 shows on the one hand 
that both procedures to obtain the topological signatures
work; on the other hand these f\/igures constitute to a certain
extent a numerical check for  the expression~(\ref{topsig1}). 
Furthermore, f\/igures~2 to~5 also show how 
the separated terms, which arise from the covering isometries, 
give rise to the topological signature looked upon as the sum of 
the terms in the right-hand side of~(\ref{topsig1}).

To make explicit that the above procedure employed to extract
the topological signature of the Euclidean ($k=0$) manifold 
${\mathcal T}_4$ can be similarly applied to any other classes
of 3-manifolds of constant curvature, we have performed computer-aided 
simulations for an elliptic ($k=1$) as well as a hyperbolic ($k=-1$)
manifolds. In the remainder of this section we shall brief\/ly
report the results of our simulations without going into details 
of the calculations (and programs) for the sake of brevity.

\begin{sloppypar}
We recall, f\/irstly, that all (locally homogeneous) manifolds 
with $k=1$ are known~\cite{Wolf}, and that they are the simply 
connected compact covering manifold $\widetilde{M}=S^3$, the 
(compact) projective space $P^3=S^3/Z_2$
and the following (compact) quotient manifolds $M = S^3/\Gamma$, 
where $\Gamma$ is one of the following groups: 
(i) the cyclic group of $Z_p$ of order $p>2$; 
(ii) the binary dihedral groups $D_r^{*}$ of order $4r$ ($r\geq 2$); 
(iii) the binary polyhedral groups: $T^*$ (where $T$ is the symmetry
group of the regular tetrahedron), $O^*$ (where $O$ is the symmetry
group of the regular octahedron), and $I^*$ (where $I$ is the symmetry
group of the regular icosahedron);
(iv) the groups of the form $D_r^* \times Z_p$ with $p \geq2$ and 
$r \geq 2$, although for certain values of $r$, $D_r^* \times Z_p$ 
can act on $S^3$ in two different ways; and 
(v) the groups of the form $H \times Z_p$ 
($p \geq 2$), where $H = T^*, O^*, I^*, T_n^*$, and the groups 
$T_n^*$ are non-cyclic subgroups of $T^*$.
\end{sloppypar}

We have performed computer simulations for the specif\/ic elliptic 
3-manifold $S^3/Z_5\,$, whose volume $2\,\pi^2\,R^3/5\,$ is one f\/ifth 
of the volume of $S^3$. A FP (tetrahedron) together with the pairwise 
faces identif\/ications is given by Weeks~\cite{Weeks85}. We have
taken as the observed universe the whole covering space $S^3$, i.e.
a solid sphere with radius $a=\pi$ [$\,R=1\,$ in (\ref{RWmetric})]. 
Thus all catalogues in our simulations for this manifold have the 
same number of images. \  
Figure~6 shows the graph of the topological signature \ 
$\varphi^S(s_i)\,=\,(n -1)\,[\,<\!\Phi(s_i)\!>-\,\,\Phi^{sc}_{exp}(s_i)\,]$
for a manifold $S^3/Z_5$ with the edge of the tetrahedron $L \simeq 1.82$
and for $m=180$, $n=100$ images ($n_s=20$), $K=3000$ 
catalogues, and where the expression~(\ref{fdensEL}) for 
$\Phi^{sc}_{exp}(s_i)$ has been used. Clearly the existence of 
topological spikes in PSH gives rise to spikes in $\varphi^S(s_i)$.
Now since the observed universe is the whole unitary sphere $S^3$ and
the graph of $\varphi^S(s_i)$ shown in f\/igure~6 presents
no spike, it becomes apparent that the covering group $\Gamma$ of this
3-manifold $S^3/Z_5\,$ has no translations.

\begin{sloppypar}
Contrarily to the case of elliptic space-forms a complete 
classif\/ication of the 3-dimensional hyperbolic manifolds has not 
yet been accomplished. However, from a topological viewpoint the
negatively curved (hyperbolic) universes are generic in the sense
that most 3-dimensional manifolds can be viewed as homogeneous
negatively curved and compact~\cite{Thurston97}.
The volume of the covering manifold $H^3$ is inf\/inite, but 
the quotient manifolds $H^3/\Gamma$ can either be f\/inite of 
inf\/inite.
\end{sloppypar}

We have performed computer simulations for the specif\/ic compact
hyperbolic 3-man\-i\-fold known as Seifert-Weber dodecahedral space, 
which is obtained by identifying or glueing the opposite pentagonal 
faces of a dodecahedron after a rotation of $3\,\pi/5\,$.  
Figure~7 shows the graph of the topological signature $\varphi^S(s_i)=
(<\!n\!> -\,\, 1\,)\,[<\!\Phi(s_i)\!>-\,\Phi^{sc}_{exp}(s_i)]$ 
for this hyperbolic space  where the centre of the 
dodecahedron was taken to coincide with the 
centre of the observed universe $\mathcal{B}_a\,$, whose diameter 
is $2a= 2.88$. The length $L$ of the edges of the pentagonal faces 
and the height $H$ of the dodecahedron are such that 
$L=H=1.992769\,$, where the lengths are measured with the 
hyperbolic geometry~(\ref{RWmetric}) with $R=1$. 
We have taken $m=100\,$ bins, $n_s=10$ seeds, ($<\!n\!>\, \simeq 18$) 
$K=16000$ catalogues, and used the expression~(\ref{fdensH}) for 
$\Phi^{sc}_{exp}(s_i)$. Since the existence of topological spikes 
in PSH gives rise to spikes in $\varphi^S(s_i)$ and reciprocally,
f\/igure~7 also shows that the PSH of even this highly symmetrical 
hyperbolic manifold presents no topological spikes, clarifying the
conjecture made in~\cite{FagGaus} and
in agreement with the results of~\cite{GTRB98}.

It should be stressed that the unit of
lengths used in all simulations of this section could certainly
have been taken to be of hundreds or thousands of $Mpc$
obviously without changing the patterns of the simulations we
have performed. We have avoided large numbers for the sake 
of simplicity.

To close this section we mention, for the sake of completeness, 
that small multiply-connected f\/lat models (like those
with $\mathcal{T}_4$ topology) may be consistent 
with COBE CMB constraints (this is still a controversial point,
see in this regards, e.g., Levin \emph{et al.\/}~\cite{LSS98},
Roukema~\cite{Roukema2000} and Inoue~\cite{Inoue2000}).
Nevertheless, according to the current values of the cosmological
parameters, a spherical universe is unlikely to be completely 
observable. Indeed, from the redshift-distance relation 
for Friedmann-Lema\^{\i}tre-Robertson-Walker with dust and
cosmological constant, one can easily obtain that a spherical 
universe with matter current density parameter $\Omega_{m0}=0.3$ 
would require that $\Omega_{\Lambda 0} > 1.35$ for the antipodal 
point to be visible at a redshift below $z=1000$, i.e.  for $S^3$ 
to be inside the sphere representing the microwave background. 
Similarly with $\Omega_{m0}=0.3$ would require that 
$\Omega_{\Lambda 0} >1.62$ for an antipodal point to be visible
at a redshift below $z=3$, i.e. for $S^3$ to be
coverable by galaxies and quasars.
These range of values for $\Omega_{\Lambda 0}$ are
in conf\/lict with current constraints on $\Omega_{\Lambda 0}$ 
and on $\Omega=\Omega_{m0} +\Omega_{\Lambda 0}$ 
(see, for example, Lange \emph{et al.\/}~\cite{Lange01},
Bond \emph{et al.\/}~\cite{Bond00}, Balbi \emph{et al.\/}%
~\cite{Balbi00} and Bernardis \emph{et al.\/}~\cite{Bernardis00}).

\vspace{5mm}
\section{Concluding Remarks}
\label{remarks}
\setcounter{equation}{0}

If we live in a multiply connected Friedmann-Lema\^{\i}tre-%
Robertson-Walker (FLRW) universe the sky may show multiple 
correlated images of cosmic objects. The image correlations 
are dictated by the discrete isometries of the covering 
group $\Gamma$ of the 3-manifold used to model its space 
sections. The periodic distribution of images gives rise 
to correlations in their positions. In the crystallographic 
method~\cite{LeLaLu} these correlations are couched 
in terms of correlations in distances between the 
images. Actually, the method of cosmic crystallography (CC) 
looks for distance correlations between cosmic images using 
pair separations histograms (PSH), with normalized function 
given by~(\ref{defpsh}), and whose graph gives the (normalized)
number of pairs of sources versus the distance between them.

The primary expectations in CC were that the distance correlations 
would manifest as topological spikes in PSH's, and that the spike 
spectrum would be a def\/inite signature of the topology.
Although the f\/irst simulations performed for specif\/ic 
f\/lat manifolds appeared to conf\/irm the initial 
expectations~\cite{LeLaLu}, histograms generated subsequently 
for hyperbolic manifolds~\cite{LeLuUz,FagGaus} revealed that 
the PSH's of those manifolds exhibit no spikes.
Concomitantly, a theoretical statistical analysis of the 
distance correlations in the PSH's was performed, and a formal 
proof was presented that the spikes of topological origin
in PSH's are due to translations alone~\cite{GTRB98}.
This result explains the absence of spikes in the PSH's of 
hyperbolic manifolds, and also gives rise to the fact that
Euclidean distinct manifolds which admit the same translations
on their covering group present the same topological spike 
spectrum~\cite{GTRB98,GRT99}.

Although the set of topological spikes in PSH's is not def\/inite 
topological signature and is not suf\/f\/icient for distinguishing 
even between some compact f\/lat manifolds~\cite{GRT99}, the most 
striking evidence of multiply-connectedness in PSH's is indeed the 
presence of topological spikes, which arise 
only when the isometry is a Clif\/ford translation. 
The other  isometries, however, manifest as tiny deformations 
of the expected pair separation histogram (EPSH) corresponding 
to the underlying simply connected universe. 
In computer-aided simulations, however,  histograms contain 
statistical f\/luctuations, which can give rise to sharp 
peaks of statistical origin, or can hide (or mask) the tiny 
deformations due to non-translational isometries. 

The most immediate approach to cope with f\/luctuation problems 
is through the reduction of the noise in PSH's by using the 
mean pair separation histogram (MPSH) scheme to obtain 
$<\!\Phi(s_i)\!>$  rather than a single PSH $\,\Phi(s_i)\,$. 
However, in most of the computer simulations we have performed,
for a reasonable number of images ($n \simeq 60 \,$, or so)
and when there is no topological spikes (no translations), 
the graphs of the $\,\Phi_{exp}(s_i)\, \simeq \,\,<\!\Phi(s_i)\!>\,$ 
and $\,\Phi^{sc}_{exp}(s_i)\,$ are essentially the same (the 
deformation of the EPSH due to the non-translational isometries
are indeed rather tiny!),
making clear that in practice $\,\Phi_{exp}(s_i)$ is not
a suitable quantity for revealing the topology of multiply 
connected universes.

In this work we have studied two ways of obtaining 
(extract and construct) the topological signature of any multiply 
connected RW universe. The important points in the f\/irst 
approach are: (i) the use of~(\ref{EPSH4}) to introduce a 
\emph{new} quantity $\varphi^S(s_i) \equiv (n-1)\,[\Phi_{exp}(s_i) 
- \Phi^{sc}_{exp}(s_i)]\,$, which turns out to be suitable for 
revealing the topological signature; (ii) the supplementary use 
the MPSH technique to drastically reduce the statistical noises 
and therefore improve the approximations 
$<\!\Phi(s_i)\!> \,\simeq  \Phi_{exp}(s_i)\,$ and
$<\!\Phi^{sc}(s_i)\!> \,\simeq \Phi^{sc}_{exp}(s_i)\,$.%
\footnote{As a matter of fact, although in general necessary, 
in the present article we did not use the approximation
$<\!\Phi^{sc}(s_i)\!> \,\simeq \Phi^{sc}_{exp}(s_i)\,$, but the
explicit expressions for $\Phi^{sc}_{exp}(s_i)$ given by~%
(\ref{fdensE})~--~(\ref{fdensEL}) for universes 
fulf\/illed with an uniform distribution of 
cosmic objects. }

\begin{sloppypar}
The second way of obtaining  the topological signature of a 
multiply connected RW universe (constructive approach) is based 
on the explicit expression for $\varphi(s_i)$ given by~%
(\ref{topsig1})~--~(\ref{PhiGamma}). In this approach the 
topological signature is obtained by considering the sum of 
the terms on the right-hand side of~(\ref{topsig1}) together 
with the MPSH technique to reduce the statistical noises.
\end{sloppypar}

We have also shown through concrete computer-aided simulations 
and based upon our theoretical results, that the two ways of 
obtaining the topological signatures of multiply connected RW
manifolds give rise to the same topological signature, as one 
would have expected from the outset. Furthermore, the strength 
of these approaches has been shown by extracting the topological 
signatures of a f\/lat ($k=0$), an elliptic ($k=1$), and a 
hyperbolic ($k=-1$) multiply connected RW universes. 
There emerges from our theoretical results and simulations that 
the topological signature arises (in the mean) even when there 
are just a few images for each object.

It should be emphasized that although in the simulations of 
section~\ref{simula} we have assumed that the matter is 
homogeneously distributed in the universe, the technique we 
have discussed in this article can be similarly used for
other types of matter distribution. In the more realistic
cases in which the galaxies (or cluster of galaxies) cluster, 
the matter distribution is clearly inhomogeneous. This inhomogeneity 
has been cast in terms of statistical indicators such as correlation 
functions, which can be used either for the computation of the 
expression for $\Phi^{sc}_{exp}(s_i)$ or to simulate the clustered 
distribution of matter in the simply-connected case to have the 
approximate contribution $\sim  \Phi^{sc}_{exp}(s_i)$. Once either 
of these steps has been carried out one can proceed along the 
lines described in this paper.

It should be stressed that the ultimate 
step in most of the statistical approaches to extract the 
topological signature is the comparison of the signature
obtained from simulated catalogues against similar ones generated 
{}from real catalogues. To do so one clearly has to have the simulated 
patterns of the topological signatures of the manifolds, which can
be achieved by the approaches discussed in this paper. Note, however,
that the MPSH technique (needed in both approaches) is restricted to 
simulated catalogues, since it is impossible in practice to have 
a reasonable ensemble of comparable catalogues of real cosmic sources 
to calculate mean and expected values of the relevant quantities.
An approach to face the important remaining f\/luctuations 
problem is to study quantitatively the noise which arises 
in~(\ref{nonpropquant}) in order to develop suitable 
f\/ilters for those statistical noises which naturally arise in 
PSH's built from real catalogues.

Note, however, that we have assumed that all cosmic objects of 
our interest are pointlike and have long lifetimes so that none 
was born or dead since the time corresponding to the redshift
cutoff of the catalogue. Moreover, we have also assumed that 
all objects are comoving, so that their worldlines have constant 
spatial coordinates. Although simpliflying these assumptions are 
commonly used in the literature and are very useful to study the 
observational consequences of a non-trivial topology for the 
universe.

The main sources of uncertainties of the known statistical
approaches to determine the topology of our universe from discrete
sources are (see, for example,~\cite{GTRB98} and~\cite{LeUzLu}):
\begin{enumerate}
\item[(i)]
the cosmological parameters $\Omega_{m0}$ and $\Omega_{\Lambda 0}$
are not accurately known or determined, so one cannot accurately 
compute distances from redshif\/ts. The uncertainty in the 
determination of the cosmological parameters give rises to uncertainty 
in the determinations of the positons of the sources, which in turn 
lead to errors in the determination of distances between pairs of 
sources.
\item[(ii)]
the cosmic objects are excactly not comoving, so multiple images are 
not where they ought to be: in real catalogues there are 
uncertainties in the positions of the objects due to their peculiar 
velocities.
\item[(iii)]
most cosmic objects do not have very long lifetimes, so there 
may not exist images of the same object at very large dif\/ferent 
distances from one of our images (observer).
\item[(iv)] 
in most studies of such statistical approaches to cosmic 
topology the catalogues are taken to be complete. Real catalogues,
however, are incomplete: objects are missing due to selection
rules, and also most surveys are not full sky coverage 
surveys.
\end{enumerate}

It should be noticed that if  the universe turns out to be small 
with a covering group containing Clif\/ford translations, the 
collecting correlated pairs method (CCP method), 
which has been recently devised by Uzan, Lehoucq and
Luminet~\cite{UzLeLu99} (see also~\cite{LeUzLu}), can be used 
in conjunction with the cosmic crystallographic method to 
determine both the cosmological parameters $\,\Omega_{m0}$ and 
$\,\Omega_{\Lambda 0}$ as well as the topology of the universe. 
In the remainder of this paper we shall discuss this point in 
more details.

Instead of grouping the pairs of sources according to their
separations to extract the signature of the correlated pairs, 
in the CCP method one uses the basic feature of the isometries, 
 i.e., that they preserve the distances between pairs of images
regardless of whether they are correlated or not.
Thus, if $(p,q)$ is a pair of arbitrary images (correlated or not)
in a given catalogue $\mathcal{C}$, then for each $g \in
\Gamma$ such that the pair $(gp,gq)$ is also in $\mathcal{C}$
we obviously have%
\footnote{The pairs for which eq.~(\ref{typeI}) holds have been 
referred to by Uzan \emph{et al.}~\cite{UzLeLu99} as 
\emph{type I pairs}.}
\begin{equation}   \label{typeI}
d(p,q) = d(gp,gq) \; .
\end{equation}
This means that for a given (arbitrary) pair $(p,q)$ of images 
in $\mathcal{C}$, if there are $n$ isometries $g \in \Gamma$ such 
that both images $gp$ and $gq$ are still in $\mathcal{C}$, then
the distance $d(p,q)$ will occur $n$ times.%

The CCP method consists essentially in a count of the number
of times each pair distance between images is repeated in a given
catalogue.%
\footnote{Note that one necessarily has a nonnull CCP-index
when there are pattern repetitions in a catalogue $\mathcal{C}$ 
as that of the scheme used by Roukema~\cite{Roukema96} in the 
search for repetition of local quintuplets of quasars in our 
universe. As he has pointed out his choice for quintuplet 
rather than any $n$-tuples was a compromise between the data 
and computing power available, but the isometries clearly
preserve the local $n$-tuplets.} 
Actually, the CCP-index $\mathcal{R}$ is def\/ined to 
be the quotient between the result of this count and the total 
number of pair of sources in $\mathcal{C}$ minus $1\,$, i.e. $(N-1)$ 
in the notation of the second Section.

Now, given that in a catalogue corresponding to a simply 
connected universe the probability that a given pair separation 
be repeated is zero, a non-zero CCP-index is an indicator 
of a non-trivial topology (multiply-connectedness) of our Universe. 
However, as it has been discussed in~\cite{UzLeLu99}, the CCP-index 
$\mathcal{R}$ gives no hint on what is the topology of the spatial 
sections of the universe.

It should be noticed that cosmic crystallography, its variant 
approach for simulated catalogues discussed in this work, as well 
as the CCP method, are them all based on the distance determination 
between cosmic images, and therefore they are rather sensitive 
to the (precise) values of the cosmological parameters 
$\Omega_{m0}$ and $\Omega_{\Lambda 0}$. 
However, as it has been discussed in ref.~\cite{UzLeLu99}, in a 
small universe the graph of $\mathcal{R}$ in terms of the 
cosmological parameters $\,\Omega_{m0}$ and $\,\Omega_{\Lambda 0}\,$ 
should exhibit a resonance spike at the correct values of these
parameters. Therefore, one can use the CCP method to achieve a 
precise determination of these parameters. So, if the Universe 
turns out to be (small) Euclidean or spherical, for example, 
one can use the values of $\Omega_{m0}$ and $\Omega_{\Lambda 0}$ 
determined through the CCP method, and then apply cosmic 
crystallography (CC) to determine the spike spectrum in the PSH. 
If the spike spectrum does not uniquely determine the topology 
of our Universe (as in the case for compact Euclidean manifolds), 
one can use again the CCP-index in order to distinguish among
the possible candidates. 
Indeed, since the CCP-index is essentially a multiple count 
of the number of isometries that give rise to repeated 
distances of images in the observed universe, then the CCP-index
corresponding to a given covering group is larger than the one
corresponding to any of its subgroups. Therefore, detailed 
simulations of the CCP method applied to each likely candidate
(manifold) can eventually permit to single out one of 
the possible manifolds (candidates), determining therefore 
the topology of the universe. This combined method (CCP plus 
CC and/or its variant approach for simulated catalogues)
is currently under investigation, and our results will be 
shortly published elsewhere.

\vspace{0.5cm}
\section*{Captions for the f\/igures}
\begin{description}
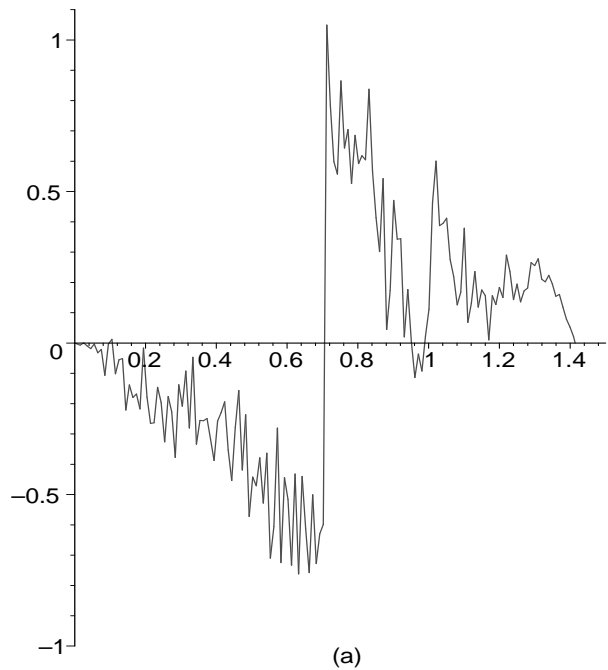
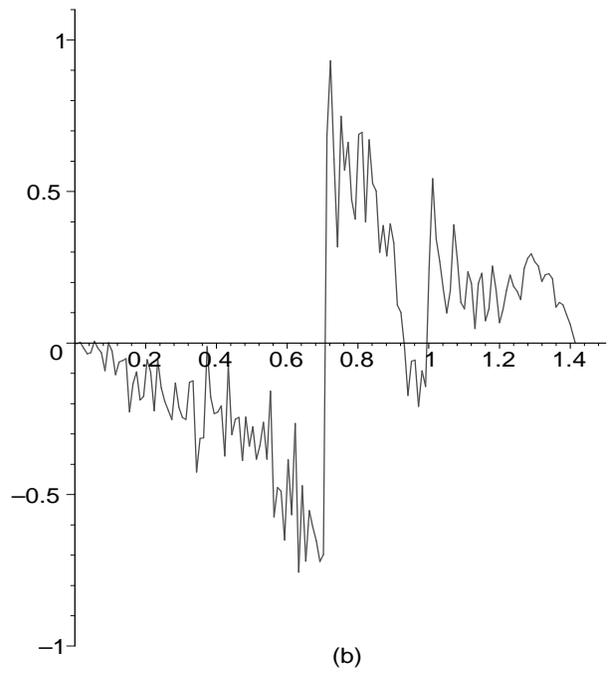
\item[Figure 1.] The topological signature $\varphi^S(s_i)=
(<\!n\!> -\,1\,)\,[<\!\Phi(s_i)\!> - \,\Phi^{sc}_{exp}(s_i)]$ 
(obtained through the MPSH technique) of the cubic manifold 
${\mathcal T}_4$ with $L=1$, in an observed universe with 
radius $a \simeq 0.71\,$.
The horizontal axis gives the pair separation $s$ while the
vertical axis furnishes the normalized number of pairs.
In (a) the number of seeds is $n_s=15$ and corresponds to an 
average number of images per catalogue $<\!n\!> \,\,\simeq 23\,$.
In (b) the number of seeds is $n_s=100$ and corresponds to an 
average number of images per catalogue $<\!n\!>\,\, \simeq 153\,$.
In both cases one arrives at essentially the same topological
signature.

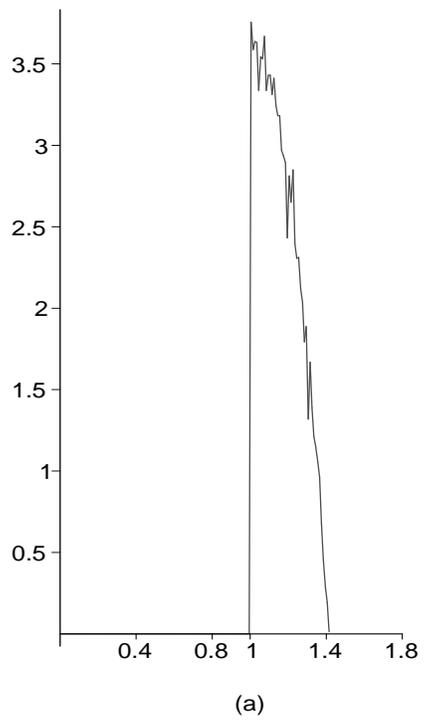
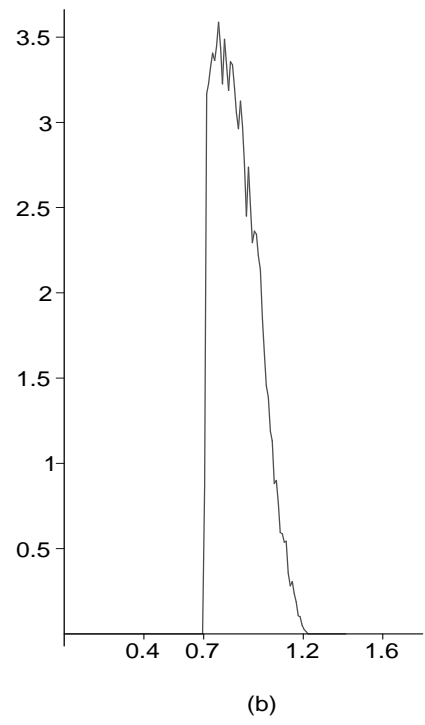
\item[Figure 2.] The contribution $\Phi^g(s_i)$ of the isometries
$g=\alpha$ [part (a)] and $g=\beta$ [part (b)] to the topological
signature $\varphi^S(s_i)$ of the cubic manifold ${\mathcal T}_4$ 
with $L=1$ in the observed universe with radius $a \simeq 0.71\,$.
The horizontal axis gives the pair separation $s$ while the
vertical axis gives the normalized number of pairs.
The MPSH approach was used to obtain both MPSH's $<\!\Phi^g(s_i)\!>\,$. 
These f\/igures make clear that the action of the isometries
$\alpha$ and $\beta$ starts, respectively, at $s=1$ and $s \simeq 0.71$.
These starting points for the actions of those isometries give
rise to the two discontinuities of the topological signatures
$\varphi^S(s_i)$ of f\/igures~1a and 1b. 

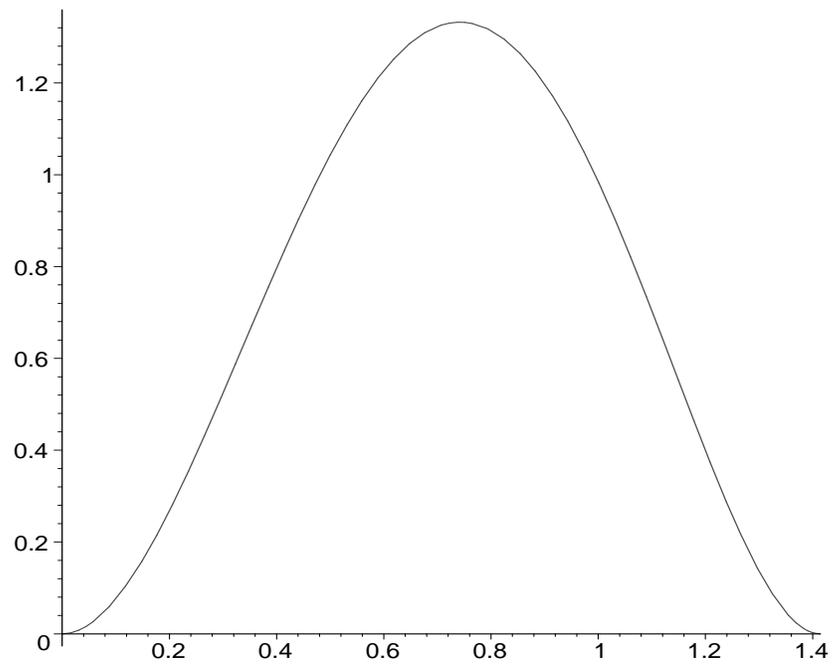
\item[Figure 3.] The EPSH $\Phi^{sc}_{exp}(s_i)$ as given by
eq.~(\ref{fdensE}) for an Euclidean universe $\mathcal{B}_a$
with radius $a\simeq 0.71\,$ and fulf\/illed with an uniform 
distribution of cosmic objects.
The horizontal axis gives the pair separation $s$ while the
vertical axis gives the normalized number of pairs.

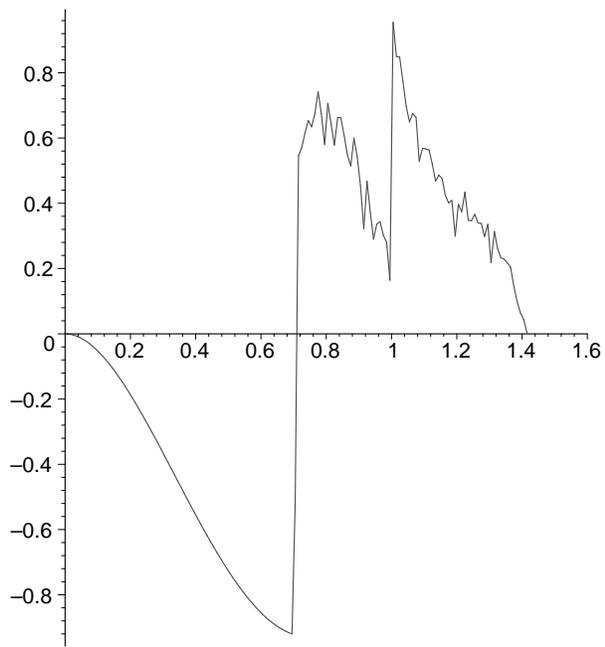
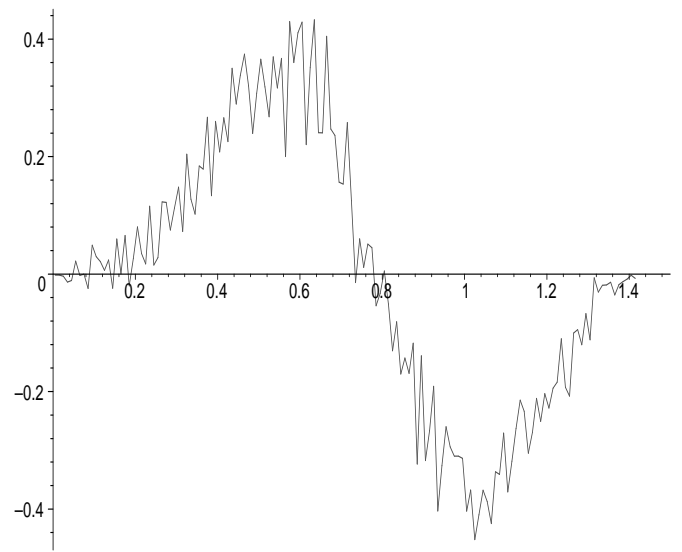
\item[Figure 4.] The contribution of the terms $\varphi^{\Gamma}(s_i)$
[part (a)] and $\varphi^U(s_i)$ [part (b)] as given, respectively, by 
eqs.~(\ref{PhiGamma}) [or (\ref{PhGama})] and (\ref{PhiU}), to the 
topological signature $\varphi^S(s_i)$ of the cubic 
manifold ${\mathcal T}_4$ with $L=1$ in the observed universe 
with radius $a \simeq 0.71\,$. 
The horizontal axis gives the pair separation $s$ while the
vertical axis gives the normalized number of pairs.
The exact expression~(\ref{fdensE}) was used in part~(a) and 
MPSH technique was used in both part~(a) and part~(b). 

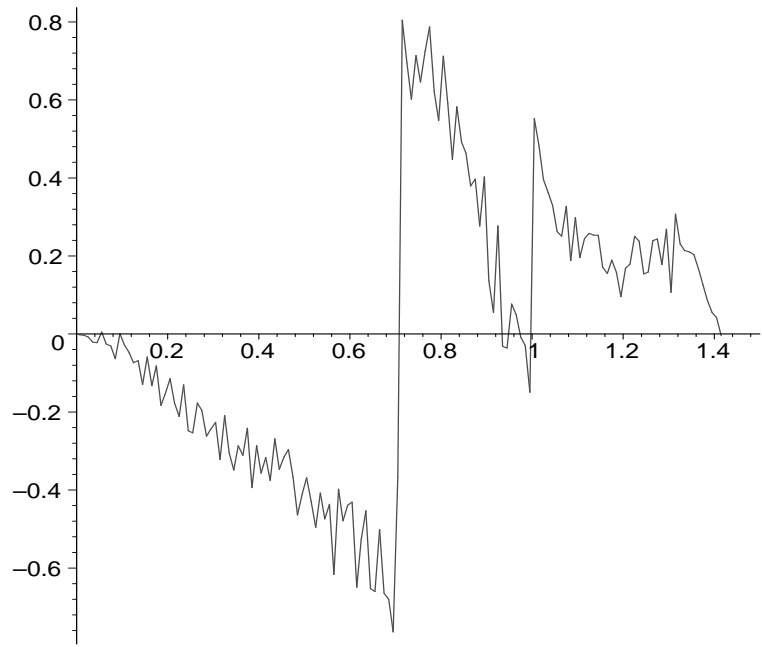
\item[Figure 5.] The sum of the contributions due to the terms 
$\varphi^{\Gamma}(s_i)$ and $\varphi^U(s_i)$, given in f\/igure~4,
to the topological signature $\varphi^S(s_i)$ [as given by 
eqs.~(\ref{topsig1})~--~(\ref{PhiGamma})] of the cubic manifold 
${\mathcal T}_4$ with $L=1$ in the observed universe with radius 
$a \simeq 0.71\,$. 
The horizontal axis gives the pair separation $s$ while the
vertical axis furnishes the normalized number of pairs.
The comparison of this f\/igure with f\/igure~1 shows that the
two procedures for obtaining the topological signatures give 
rise to the same topological signature.

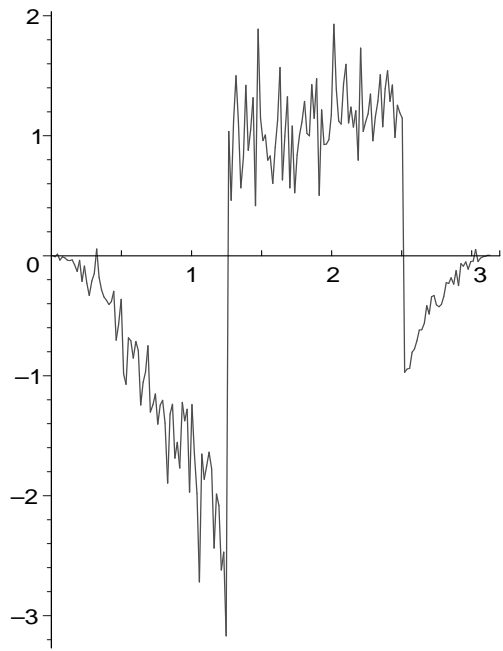
\item[Figure 6.] The topological signature $\varphi^S(s_i)=
(n -1)\,[<\!\Phi(s_i)\!>-\,\Phi^{sc}_{exp}(s_i)]$ for an elliptic 
manifold $S^3/Z_5$ with the edge of the tetrahedron $L \simeq 1.82$.
The observed universe was taken to be the whole unitary sphere $S^3$,
and the expression~(\ref{fdensEL}) for $\Phi^{sc}_{exp}(s_i)$ 
was used. The horizontal axis gives the pair separation $s$ while the
vertical axis gives the normalized number of pairs.

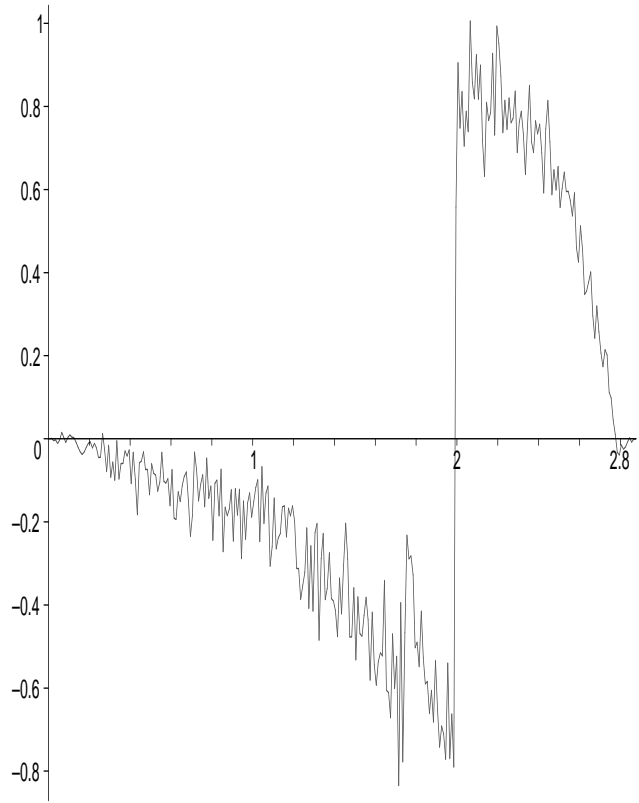
\item[Figure 7.] The topological signature $\varphi^S(s_i)=
(<\!n\!> -\,\, 1\,)\,[<\!\Phi(s_i)\!>-\,\Phi^{sc}_{exp}(s_i)]$ 
for the Seifert-Weber dodecahedral (hyperbolic) space  with 
$L=H=1.992769\,$ (edges $L$ and height $H$), where the lengths 
are measured with the hyperbolic geometry~(\ref{RWmetric}) with 
$R=1$. The observed universe $\mathcal{B}_a\,$ with diameter 
$2a= 2.88$ was used. The expression~(\ref{fdensH}) for 
$\Phi^{sc}_{exp}(s_i)$ was used.
The horizontal axis gives the pair separation $s$ while the
vertical axis furnishes the normalized number of pairs. 
\end{description}

\end{document}